\documentclass[final,5p,times,twocolumn,lefttitle]{elsarticle}
\usepackage{amssymb}
\usepackage{lipsum}
\usepackage{amsmath}
\usepackage{xcolor}
\usepackage{color}
\usepackage{float} 
\usepackage{graphicx}
\usepackage{caption}
\usepackage{subcaption}
\usepackage{soul}
\usepackage[T1]{fontenc} 
\setcitestyle{square} 
\biboptions{comma}
\usepackage[colorlinks=true, linkcolor=red, citecolor=green, urlcolor=magenta]{hyperref}

\begin{document}

\begin{frontmatter}
\title{Generalized ModMax and the Early Universe}

\author[First,Second]{M. Sabido}
\ead{msabido@fisica.ugto.mx}
\author[Third]{V. Sierra}
\ead{vicente.sierra@ligo.org}
\affiliation[First]{organization={Departamento de F\'isica de la Universidad de Guanajuato},
addressline={A.P. E-143}, city={Le\'on},    
            postcode={37150},
            state={Guanajuato},
            country={M\'exico}} 
\affiliation[Second]{organization={Department of Theoretical Physics, University of the Basque Country UPV/EHU},
addressline={P.O BOX 644}, city={},
            postcode={48080}, 
            state={Bilbao},
            country={Spain}}
\affiliation[Third]{organization={Departamento de F\'isica, Universidad de Guadalajara},
addressline={}, city={Guadalajara},
            postcode={44430}, 
            state={Jalisco},
            country={M\'exico}}
\begin{abstract}

In this  work, we study a cosmological model driven by Generalized ModMax nonlinear electrodynamics. We find that, with an appropriate choice of the theory's parameters, the universe's initial singularity can be avoided. Moreover, we also find that this model has an inflationary epoch that is consistent with the current values for $N$, $n_s$ and $r$. Therefore, using Generalized ModMax, we can construct a non singular Universe with an inflationary epoch.
\end{abstract}
\end{frontmatter}

\section{Introduction}\label{introduction}
The standard model of cosmology provides a successful framework for understanding the history and evolution of the Universe. Nonetheless, it is essentially incomplete as it requires very specific initial conditions. It postulates a uniform cosmological background, described by a spatially-flat, homogeneous and isotropic Friedmann-Robertson-Walker (FRW) metric. Within this setting, it also requires an initial, almost scale invariant distribution of primordial density perturbations and a nearly flat space-time. The best solution for this complexities is the mechanism known as inflation. Inflation postulates a period of accelerated expansion in the very early Universe, preceding the radiation era, which offers a physical model for the origin of these initial conditions. Such period of accelerated expansion drives a curved FRW spacetime towards spatial flatness, and expands the causal horizon beyond the present Hubble length while containing all the relevant scales to describe the large-scale structure.

In order to induce an accelerating scale factor, a negative pressure in the energy-momentum tensor is needed. The simplest way to achieve this is to introduce a constant term  (i.e. a cosmological constant as in dark energy), but this is discarded as this would cause never-ending inflation. The most common approach is to use a scalar field known as the \textit{inflaton}. This time dependent scalar field, in the slow-roll regime, can induce the inflationary epoch with a graceful exit to the classical evolution. Of course, there are other models that induce an inflationary epoch, i.e. Starobisnki inflation, chaotic inflation, string theory inspired models (for a current review see \cite{Kallosh:2025ijd}), and others. 

Another drawback of the standard cosmological model is the initial singularity. However, it has been proven that this singularity can be avoided by introducing nonlinear electrodynamics (NLED) as the main energetic component of the early Universe. Moreover, it is known that one can induce an accelerating scale factor with NLED, making the study of NLED in cosmology very appealing.

The first NLED model was presented in 1934 by Born and  Infeld \cite{BI}, whose goal was to solve the electron's self-energy and finite radius problem. They proposed a Lagrangian dependent on a non-linear functional of $\mathcal{F}=\frac{1}{4} F_{\mu \nu} F^{\mu \nu}$ and $\mathcal{G}=\frac{1}{4} F_{\mu \nu} \tilde{F}^{\mu \nu}$, the Maxwell scalars, and showed that such a theory, for the electrostatic case of a point-like charge, provides a non-singular, maximal electric field at the origin, effectively achieving their goal. Ever since the introduction of the Born-Infeld electrodynamics, many other NLED models have been presented, contributing to the today wide \textit{NLED zoo} \cite{Benaoum}. Each NLED model possesses its unique properties and characteristics as, for example, the introduction of new symmetries, corrections at the strong-field limit, and the vanishing of singularities, being these last two, probably the most attractive features of non-linear electrodynamics. 
In \cite{BandosModMax}, authors showed the existence of only two non-linear electrodynamics models that possessed Maxwell's electrodynamics' symmetries: conformal symmetry and duality invariance \cite{Fushchich}. One of these models, the strong-field limit of the theory, corresponded to Bialynicki-Birula's electrodynamics \cite{BB}, which was already presented in 1983 and widely studied. The other one, which is the weak-field limit, corresponded to a new family of electrodynamics known as ModMax. A generalization of ModMax was presented in \cite{KruglovModMax}, a theory represented by a Lagrangian that has three extra parameters which, at specific values, reproduces ModMax. 

The purpose of this paper is to study Generalized ModMax in the early Universe. In particular, we focus in two aspects of cosmology: the initial singularity and the inflationary epoch. More precisely, we want to stablish if Generalized ModMax electrodynamics can simultaneously solve the initial singularity problem and induce an inflationary epoch that is consistent with the current CMB data. We find that, under specific values for the parameters of Generalized ModMax, the answer to the question is positive.
 
This paper is organized as follows. In Sec.~\ref{sec:GMM} we review ModMax and Generalized ModMax NLED. In Sec.~\ref{sec:Cosmo}, Generalized ModMax cosmology is presented and the question of the initial singularity is analyzed.  In Sec.~\ref{sec:Inflation}, we question if Generalized ModMax driven inflation is addressed. Lastly, Sec.~\ref{conclusions} is devoted to discussion and final remarks.

\section{Generalized ModMax Model} 
\label{sec:GMM}
The ModMax non-linear electrodynamics in \cite{BandosModMax} is described by the Lagrangian
\begin{equation}
    \mathcal{L}_\text{MM} = - \mathcal{F} \cosh(\gamma) + \sqrt{\mathcal{F}^2 + \mathcal{G}^2} \sinh(\gamma), \label{eq:01}
\end{equation}
where $\gamma$ is a dimensionless parameter. Some properties of this theory can be read directly from the Lagrangian as, for example, that it possesses conformal symmetry on-shell since it is of the form $\mathcal{L} = \mathcal{F} f\left(\frac{\mathcal{G}}{\mathcal{F}}\right)$; or that it recovers Maxwell's Lagrangian for $\gamma=0$. Furthermore, and more importantly, for the electrostatic case of a point-like charge, this theory yields a singularity in the electric field as $r\to 0$, such as in Maxwell's electrodynamics.

In order to smooth out singularities from the theory, in 2021, Kruglov proposed an alternative Lagrangian \cite{KruglovModMax}:
\begin{equation}\label{eq:02}
        \mathcal{L}_\text{GMM}  = \frac{1}{\beta} \left[ 1 - \left( 1- \frac{\beta \mathcal{L}_\text{MM}}{\sigma} - \frac{\beta \lambda \mathcal{G}^2}{2\sigma}\right) ^\sigma\right],
\end{equation}
where the parameters $\beta$ and $\lambda$ have dimensions of (length)$^2$, while $\sigma$ is dimensionless. This theory is named Generalized ModMax since it represents a very general NLED model. Indeed, by fixing the values of the parameters from Eq.\eqref{eq:02} other NLED models can be retrieved. For example, for $\sigma=1$ and $\beta=0$, Eq.\eqref{eq:01} is recovered. Also, for $\sigma=\frac{1}{2}$, $\gamma=0$, and $\lambda=\beta=\frac{1}{b^2}$, the Born-Infeld Lagrangian is obtained:
\begin{equation}
	\mathcal{L}_\text{GMM} =  b^2 -b^2\sqrt{1 + \frac{2}{b^2} \mathcal{F} - \frac{1}{b^4} \mathcal{G}^2}. \label{eq:03}
\end{equation}
Performing a series expansion for $\beta \mathcal{L}_\text{MM} \ll  1$ and $\beta \lambda \mathcal{G}^2 \ll 1$, in the weak-field limit, and setting $\gamma=0$, to second order in $\mathcal{G}$,  we get
\begin{equation}
	\mathcal{L}_\text{GMM} \approx -\mathcal{F} + \frac{\beta(1-\sigma)}{2 \sigma} \mathcal{F}^2 + \frac{\lambda}{2} \mathcal{G}^2, \label{eq:04}
\end{equation}
which, for $\lambda=\frac{7\beta(1-\sigma)}{2\sigma}$ and then $\lambda=\frac{4 \alpha^2}{45 m^4}$, is the Euler-Heisenberg NLED. Other models that can be derived from this Lagrangian, are Born-Infeld type models \cite{BIType}, a generalized Born-Infeld model \cite{BIGen}, exponential type models \cite{Hendi,NEDAcc}, and many others yet to be explored.

This general NLED model was explored thoroughly by Kruglov, who found that dual symmetry is only present in two cases: when $\sigma=1$ and $\lambda=0$, hence ModMax; and when $\sigma= \frac{1}{2}$ and $\lambda = \beta$, which is a Born-Infeld type model. On the other hand, conformal symmetry is present in the theory only when $\sigma=1$ and $\beta=0$, which occurs only for ModMax. Furthermore, it was proven that for $\sigma <1$, the electric field of a electrostatic point-like charge is given by

\begin{equation}
	E_\text{GMM}(r\to0) = \sqrt{\frac{2 \sigma}{\beta}} e^{-\frac{\gamma}{2}}, \label{eq:05}
\end{equation} 
which is non-singular and maximal at the origin. 

It is noted that this Generalized ModMax NLED model does not possess all of the previously enlisted properties at once; it is only by fixing the theory's parameters that these desirable characteristics take place. However, not only the possibility to consider different NLED models in the same Lagrangian is what makes this one attractive, but also the inclusion of one of the most important features of non-linear electrodynamics: avoiding singularities.

\section{Cosmology and Generalized ModMax Electrodynamics} 
\label{sec:Cosmo}

To construct a cosmological model based on a non-linear-type-radiation dominated Universe, we will follow  \cite{Benaoum}. We assume that the primordial Universe is purely magnetic and that gravity is coupled to a NLED model described by a Lagrangian of the form $\mathcal{L}_\text{NLED}= - \mathcal{F} f(\mathcal{F})$. Moreover, we also assume that the NLED behaves as a perfect fluid where the energy density and pressure are

\begin{equation}
	\begin{gathered}
		\rho = \mathcal{F} f, \\
		p = \frac{1}{3} \mathcal{F} (f+ 4 \mathcal{F} f_{\mathcal{F}}),
	\end{gathered} \label{eq:06}
\end{equation}
where $f_{\mathcal{F}} \equiv \frac{\partial f}{\partial \mathcal{F}}$.

Consequently, we rewrite the Generalized ModMax Lagrangian as 
\begin{equation}
	\begin{gathered}
		\mathcal{L}_{GMM} = - \mathcal{F} f(\mathcal{F})\;, \\
		f(\mathcal{F}) = \frac{1}{\beta} \left\lbrace \mathcal{F}^{\sigma-1}\left( \mathcal{F}^{-1} + \frac{\beta e^{-\gamma}}{\sigma}\right)^\sigma - \mathcal{F}^{-1}  \right\rbrace,
	\end{gathered} \label{eq:07}
\end{equation}
where the energy density and  pressure are given by
\begin{eqnarray}\label{eq:09}
	\rho_\text{GMM} &=& \frac{1}{\beta}\left[ \mathcal{F}^\sigma\left( \mathcal{F}^{-1}+ \frac{\beta e^{-\gamma}}{\sigma}\right)^\sigma-1\right],\\
	p_\text{GMM} &=& \frac{4\sigma\mathcal{F} (1+\rho_\text{GMM} \beta) }{3(\sigma e^{-\gamma} + \beta \mathcal{F})}- \rho_\text{GMM},\nonumber
\end{eqnarray}
where the last expression serves as the equation of state.
The continuity equation yields the evolution of the primordial magnetic field
\begin{equation}
	\dot{\mathcal{F}} = -4 H \mathcal{F} \Rightarrow \mathcal{F} = \mathcal{F}_\text{end}\left( \frac{a_\text{end}}{a}\right)^4, \label{eq:10}
\end{equation} 
this since $\mathcal{F}= \frac{1}{2} B^2$. Also, $\mathcal{F}_{end}$ and $a_{end}$ refer to the field and the scale factor evaluated at the end of inflation. In terms of the scale factor, Eq.\eqref{eq:09} is rewritten as
\begin{align} \label{eq:11}
	&\rho_\text{GMM}(a) = \frac{1}{\beta}\left\{ \mathcal{F}^\sigma_\text{end} \left[ \mathcal{F}_\text{end}^{-1}+ \frac{\beta e^{-\gamma}}{\sigma}\left( \frac{a_\text{end}}{a}\right)^4 \right]^\sigma-1\right\},\\
	&p_\text{GMM}(a)= \frac{1}{3 \beta}\left[ 3 + \frac{\mathcal{F}_\text{end}^\sigma \left( \frac{a_\text{end}}{a}\right)^{4\sigma} \left[ a^4 + \frac{\beta a_\text{end}^4 \mathcal{F}_\text{end} e^{-\gamma}}{\sigma}\right] (4\sigma-3)}{a_\text{end}^4 \mathcal{F}_\text{end}\beta + a^4 \sigma e^{\gamma}}\right].\nonumber
\end{align}
Particularly, for a ModMax NLED  radiation dominated Universe ($\sigma=1$ and $\lambda=0$) the energy density is

\begin{equation}
    p_\text{MM}(a) = \mathcal{F}_\text{end} e^{-\gamma} \left( \frac{a_\text{end}}{a}\right)^4, \label{eq:13}
\end{equation}
which is the same result (but with the multiplicative factor $e^{-\gamma}$) as the Maxwell result, where  $\rho_\text{rad} \sim a^{-4}$ and the scale factor evolves as $a_\text{MM}\sim\sqrt{t}$, which has the usual Big Bang singularity. For
the Euler-Heisenberg-type radiation universe (derived from Lagrangian Eq.\eqref{eq:04}), the energy density reduces to
\begin{equation}
    p_\text{EH}(a) = \frac{a_\text{end}^4 \mathcal{F}_\text{end} }{a^4} + \frac{a_\text{end}^8 \mathcal{F}_\text{end}^2 \beta(\sigma-1) }{2a^8 \sigma},\label{eq:14}
\end{equation}
which has a correction to $a$ compared to the Maxwell case. Although we don't have an analytic solution for the Friedmann equation, from an asymptotic analysis we can show that for $t\to0$, the scale factor  $a_\text{EH}(t) \sim t^{-5}$ and the late time evolution is given by $a_\text{EH}(t) \sim t^{-3}$, showing the the usual Big Bang singularity and the usual late time evolution.

\begin{figure}[ht]
	\centering
	\includegraphics[scale=0.35]{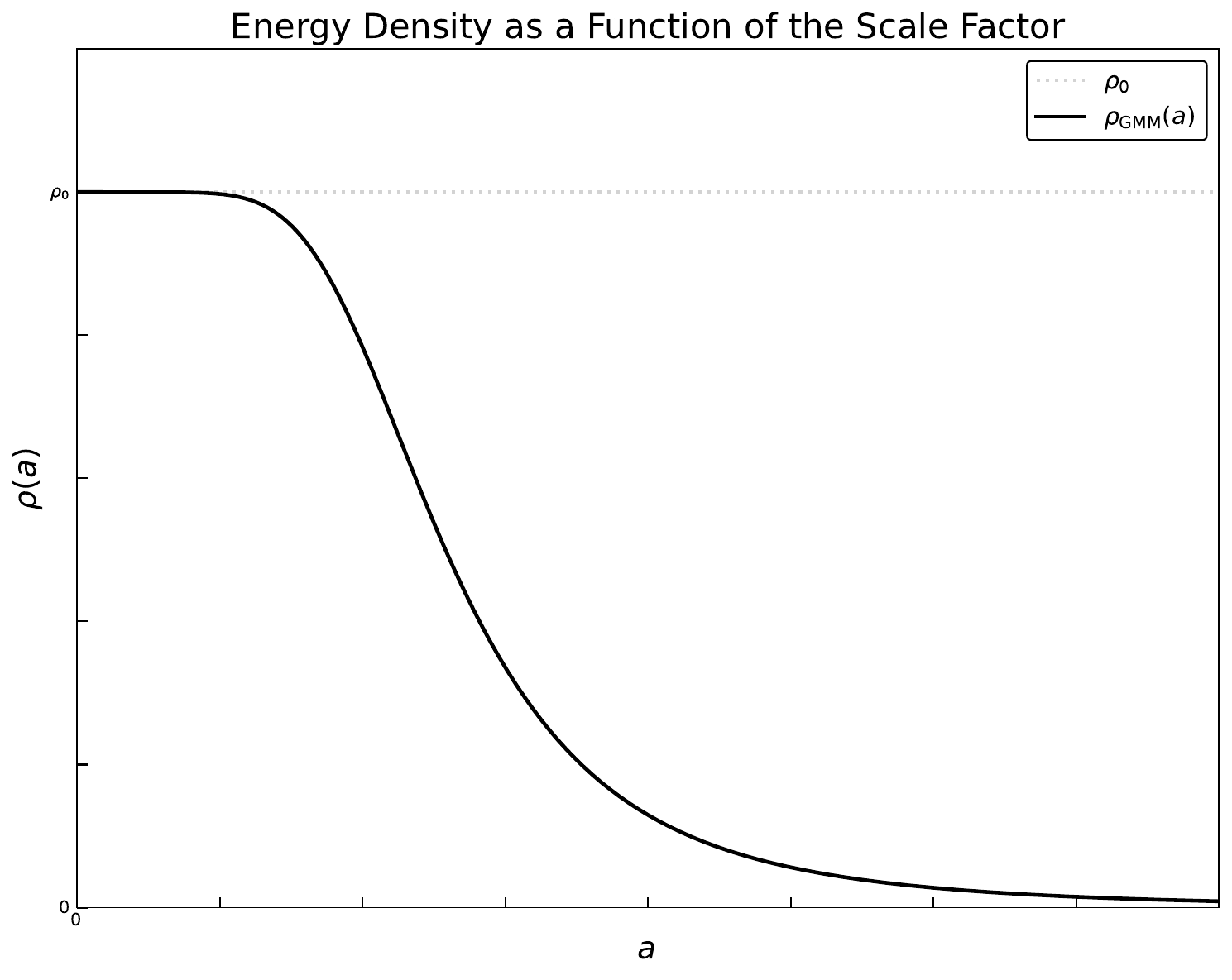}
	\caption{In a solid line, the energy density $\rho_\text{GMM}$ as a function of the scale factor $a$ for $\sigma<0$ and $\beta<0$. In a dotted line, the initial value of the energy density, $\rho_0 = \rho_\text{GMM}(a=0)$, as defined in equation \eqref{eq:16}.}
	\label{Fig1}
\end{figure}

Let us now turn our attention to the behavior of the energy density, and hence the pressure, at very early ages for the Generalized ModMax case. As it has already been mentioned, an important feature of this model is that it represents a family of NLEDs  with different properties, being of particular relevance the non-singular cases\footnote{For example, there are non-singular electric fields at the origin for $\sigma <1$.}.  
It is straightforward to show that, for $\sigma <0$, the energy density, the pressure and the curvature are regular and are given by
\begin{align}\label{eq:16}
	&\rho_0 \equiv \lim_{a \to 0} \rho_\text{GMM}(a) = - \frac{1}{\beta},\quad
    \lim_{a \to 0} p_\text{GMM}(a) = - \rho_0,\\
	&\lim_{a \to 0} R = 4 \rho_0,\quad
	\lim_{a \to 0} R_{\mu \nu} R^{\mu \nu} = 4 \rho_0^2, \quad \lim_{a \to 0} K = \frac{8 \rho_0^2}{3}.\nonumber
\end{align}
From the first expression, it might seem that the Universe could have started from a negative energy density. However, by analyzing the behavior of the energy density as a function of the scale factor for $\sigma<0$, we find that a non-pathological behavior (positive, non-diverging and continuous) occurs\footnote{This is consistent with the value needed to have a regular electric field for a point-like charge.} for $\beta <0$. In Fig.(\ref{Fig1}) we can see the behavior of $\rho_\text{GMM}(a)$ for these two conditions. 

Furthermore, the deceleration parameter \ $q = -\frac{\dot{H}}{H^2} - 1$, is
\begin{equation}
    q_\text{GMM} = \frac{2 a_\text{end} \mathcal{F}_\text{end}^{1+\sigma} \beta^3 \sigma \left( \mathcal{F}_\text{end}^{-1} + \frac{a_\text{end}^4 e^{-\gamma}\beta}{\sigma a^4} \right)^\sigma}{\left[ \mathcal{F}_\text{end}^\sigma \left( \mathcal{F}_\text{end}^{-1} + \frac{a_\text{end}^4 e^{-\gamma}\beta}{\sigma a^4} \right)^\sigma - 1\right] \left( a_\text{end}^4 \mathcal{F}_\text{end} \beta + \sigma e^\gamma a^4\right)}, \label{eq:deceleration}
\end{equation}
which, for $\sigma<0$ and $\beta<0$, results in $q<0$. 

Having established the conditions necessary on Generalized ModMax's parameters to eliminate the singularities of the energy density, we proceed to find the scale factor as a function of time and  establish its correspondence with the regularized energy density. After redefining $\sigma = - \sigma$ and $\beta = - \beta$, the first Friedmann equation is
\begin{equation}
	H^2_{GMM} = - \frac{1}{3\beta} \left[ \left( 1 + \frac{ \mathcal{F}_\text{end} a_\text{end}^4 \beta e^{-\gamma}}{\sigma}\cdot \frac{1}{a_\text{GMM}^4} \right)^{-\sigma}  -1 \right] , \label{eq:21}
\end{equation}
although we don't have analytical solutions for this, with an asymptotic analysis for small $a_\text{GMM}$, Eq.(\ref{eq:21}) simplifies to
\begin{equation}
	\dot{a}\sim  \frac{a}{\sqrt{3 \beta}}  \sqrt{1 + \left( \frac{a^4 \sigma}{ \mathcal{F}_\text{end} a_\text{end}^4 \beta e^{-\gamma}}\right)^\sigma}+ \mathcal{O}(a^{2}), \label{eq:22}
\end{equation}
and to leading order in this limit, we get the scale factor
\begin{equation}
	a(t)\sim a_{ini} e^{\frac{t}{\sqrt{3 \beta}}}, \label{eq:23}
\end{equation}
where $a_{ini}$ is the value of the scale factor in the limit 
${t\to 0}$.  On the other hand, for large $a$, to leading order, the scale factor is

\begin{equation}
	a(t)_{GMM} =\sqrt{t} \left[\sqrt{\frac{ 4\mathcal{F}_\text{end} a_\text{end}^4 e^{-\gamma}}{3}} \right ]^{1/2}, \label{eq:25}
\end{equation}
where we have set one of the integrating constants to zero. 

In a classical radiation dominated Universe, this initial scale factor is $a_{ini}=0$, which yields the usual Big Bang singularity. However, from  Eq.\eqref{eq:11} and Eq.\eqref{eq:16}, it is clear that a null initial scale factor would not make physical sense for a non-singular energy density. Moreover, in this limit we get

\begin{equation}
	\rho_\text{GMM}(a) = \frac{1}{\beta} - \frac{1}{\beta} \left( \frac{a^4 \sigma}{\mathcal{F}_\text{end} a_\text{end}^4 \beta e^{-\gamma}}\right)^\sigma + \left( \frac{a^6 \sigma}{\mathcal{F}_\text{end} a_\text{end}^4 \beta e^{-\gamma}}\right)^\sigma + \mathcal{O}(a^7), \label{eq:27}
\end{equation}

\noindent {when considering the expansion to leading order in $a$} which in conjunction with the expression for $a_{ini}$, gives

\begin{equation}
	\rho(a_\text{GMM})_{t\to0} = \frac{1}{\beta} - \frac{1}{\beta} \left( \frac{\sigma a_{ini}^4 e^{\frac{4t}{\sqrt{3 \beta}}}}{\mathcal{F}_\text{end} a_\text{end}^4 \beta e^{-\gamma}}\right)^\sigma.
\end{equation}

\begin{figure}[ht]
	\centering
	\includegraphics[scale=0.42]{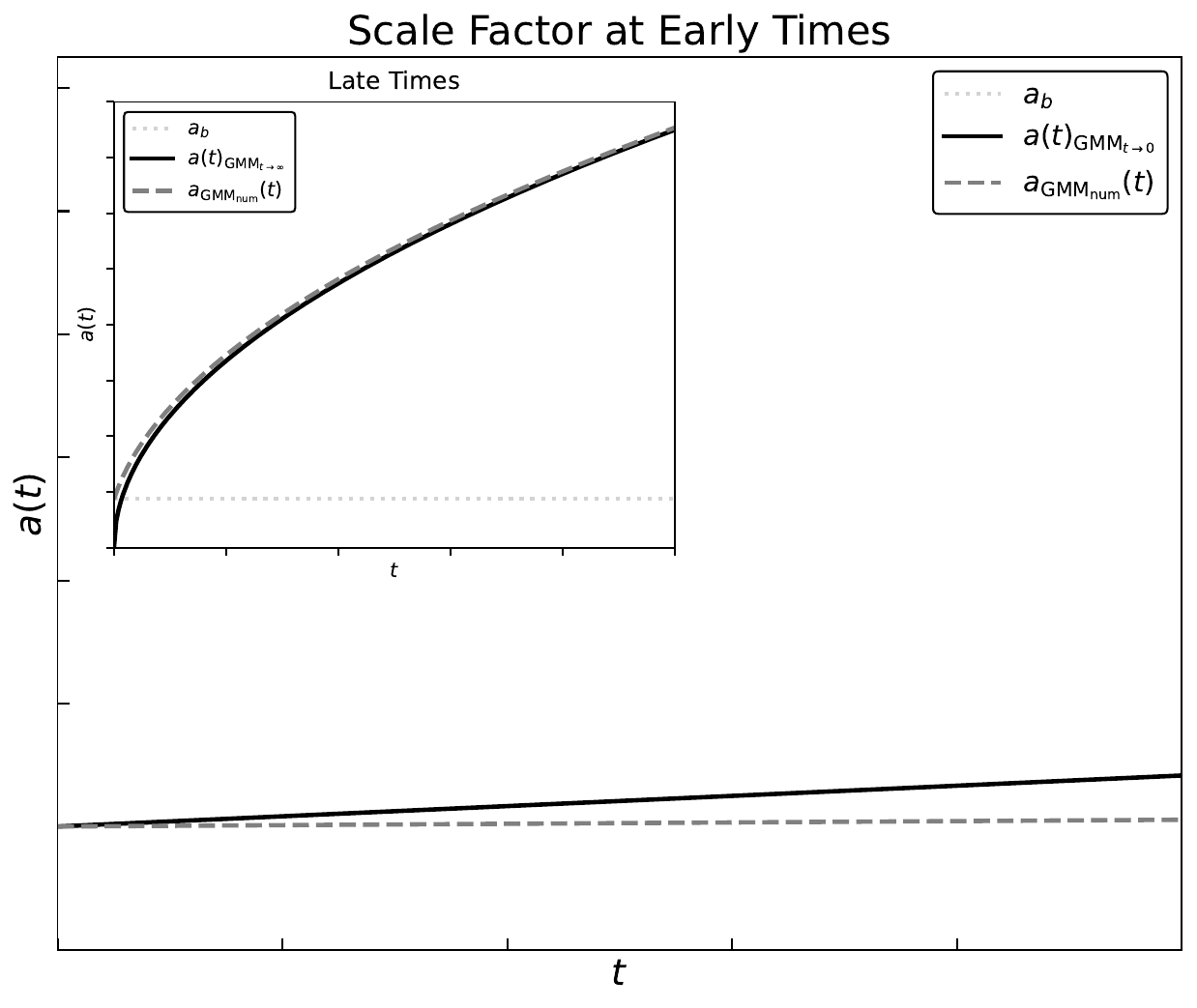}
	\caption{Scale factor as a function of time at early and late times (boxed figure). The solid line is the asymptotic behavior of $a(t)$ Eq.\eqref{eq:23} and Eq.\eqref{eq:25} respectively. The dashed plots are the numerical solutions for Eq.\eqref{eq:21}, the complete scale factor $a_\text{GMM}(t)$.}
    \label{Fig2c}
\end{figure}

\noindent Moreover, we can also obtain the initial  energy density $\rho_{ini}$ in the limit when $t\to0$:

\begin{equation}
	\rho_{ini} = \rho_0 \left[ 1 -  \left( \frac{\sigma a_{ini}^4}{\mathcal{F}_\text{end} a_\text{end}^4 \beta e^{-\gamma}}\right)^\sigma\right], \label{eq:28}
\end{equation}

\noindent which must be less than $\rho_0$ and must also correspond to that when $a_\text{GMM}=a_{ini}$. Moreover, in this limit, as the scale factor for a Generalized ModMax is given by Eq.(\ref{eq:23}),
we get 
\begin{equation}
	a_{ini} = \left( 1 - \frac{\rho_b}{\rho_0}\right)^\frac{1}{4\sigma} \left( \frac{\mathcal{F}_\text{end} a_\text{end}^4 \beta e^{-\gamma}}{\sigma}\right)^\frac{1}{4}. \label{eq:30}
\end{equation}
In Fig.(\ref{Fig2c})
, we observe the comparison between the numeric solution  of the  Friedmann equation Eq.\eqref{eq:21} and the asymptotic analysis. It is noted that the approximated solutions match the overall behavior of $a_\text{GMM}(t)$ the appropriate limit.

\section{Generalized ModMax as a Source for Inflation} 
\label{sec:Inflation}

Now that we have explored the avoidance of the singularity, we turn our attention to the question of Generalized ModMax as a source for inflation. In Eq.\eqref{eq:23}, we have the behavior of the scale factor for a Generalized ModMax radiation dominated Universe which, in the limit as $t\to0$, exhibits an exponential growth. As it can be recalled, the definition of the Universe's Inflationary period \cite{Inflation} is, essentially, a very early period of the Universe with an exponential growth. Hence, this Generalized ModMax cosmological model has a scale factor with these characteristics. This is a promising feature to consider Generalized ModMax as a possible source for an inflationary epoch. 

Let us now see if this model is consistent with the inflationary parameters \cite{Planck}
\begin{equation}
	n_s = 0.9649 \pm 0.0042 \;, \; \; 0< r < 0.064 \;, \; \; 50<N<60, \label{eq:31}
\end{equation}
{where $n_s$ is the spectral index, $r$ is the ratio of scalar to tensor perturbations and $N$ is the number of e-folds.}
As usual, the tensor-to-scalar ratio is computed as
\begin{equation}
	r = 10(2 \eta_{GMM} + 1 - n_s), \label{eq:32}
\end{equation}
where $\eta_{GMM}$ is one of the slow-roll parameters for this model. By introducing the experimental value of $n_s$, a bound for $\eta_{GMM}$ is found
and therefore the inflationary conditions on the non-linear parameters are obtained as
\begin{equation}
	\eta_\text{GMM}=\frac{2\sigma(1+\sigma)}{\mathcal{F}_\text{end} \beta e^{-\gamma} e^{4N} + \sigma} -2 \sigma < -0.01435, \label{eq:34}
\end{equation}
for $\sigma$ and $\beta >0$. This is the most general condition that can be found for the parameters so that this model can account for the Universe's Inflationary period. Nonetheless, some other necessary yet not sufficient conditions can be obtained by further analyzing Eq.\eqref{eq:34} and the possible values of some of its factors
\begin{equation}
    \beta e^{-\gamma} \mathcal{F}_\text{end} e^{4N} > 1,\quad 
        \frac{2\sigma(1+\sigma)}{\mathcal{F}_\text{end} \beta e^{-\gamma} e^{4N} + \sigma} < 2 \sigma.
    \label{eq:35}
\end{equation}
Although {these conditions are not sufficient for an inflationary epoch}, we will see that they are useful to establish values for the parameters of the model.

\section{Discussion and Final Remarks}\label{conclusions}

Until now, we have explored the cosmological implications of Generalized ModMax in cosmology, focusing our analysis on the beginning of the Universe and finding that the initial singularity can be avoided by choosing appropriate values for $\sigma$ and $\beta$.

Furthermore, in the previous section we established the general conditions in order to have an inflationary epoch. To verify if this model can satisfy the inflationary constraints, in \cite{Benaoum}, it is argued that  the value of the primordial magnetic field is $5 \times 10^{23} \; \text{cm}^{-2} \lesssim \mathcal{F}_\text{end} \lesssim 5 \times 10^{37} \; \text{cm}^{-2}$. Then, by setting $\mathcal{F}_\text{end}$, 
 we can choose $\gamma$ and $\beta$ such that, for $50<N<60$, 
the conditions in Eq.\eqref{eq:34} and Eq.\eqref{eq:35} are satisfied and,  
finally, from Eq.\eqref{eq:32}, find the range of values for $\sigma$ (for a particular number of e-folds) that is consistent with inflation. For example, taking $\mathcal{F}_\text{end} = 5 \times 10^{30} \; \text{cm}^{-2}, \; \gamma = 273, \; \beta = 22 \; \text{cm}^2$, one finds the permitted values for $\sigma$ for a particular number if {\it e-folds} as:
\begin{eqnarray}\color{red}  
    &&0.011387 < \sigma_\text{N50} < 0.018324, \\   
    &&0.006125 < \sigma_\text{N60} < 0.009825, \nonumber \label{eq:sigma_N60}
\end{eqnarray}
which are furthermore consistent with all previous discussions. 

As expected, other set of parameters can be proposed to test if the phenomenological constraints are satisfied. For example, for $\mathcal{F}_\text{end} = 5 \times 10^{30} \; \text{cm}^{-2}, \; \gamma = 350 , \; \beta = 1 \; \text{cm}^2$ and $N=50$, the first condition in Eq.\eqref{eq:35} is not satisfied, negative values for $\sigma$ which are inconsistent with the conditions for a non-singular Universe. Showing that the conditions in Eq.\eqref{eq:35} are indeed necessary. Lastly, if the following set of parameters is considered, $\mathcal{F}_\text{end} = 5 \times 10^{30} \; \text{cm}^{-2}, \; \gamma = 274, \; \beta = 27.5  \; \text{cm}^2$. For $N=50$ all conditions are satisfied; however the first condition in Eq.\eqref{eq:35} is barely satisfied, as $\beta e^{-\gamma} \mathcal{F}_\text{end} e^{4N}=1.0018$. This, in turn, results in $\sigma$ being negative suggesting that the conditions Eq.\eqref{eq:35}, although useful, are not sufficient. 

In summary, in this work we have presented a cosmological model coupled to Generalized ModMax NLED, which,  with the appropriate values for the parameters of the theory, is consistent with the inflationary constraints and simultaneously has a non singular scale factor at $t=0$. Although one can conclude that Generalized ModMax driven inflation is consistent with the phenomenological bounds on the inflationary parameters, more phenomenology is needed to narrow down all of their values.

\section*{Acknowledgments}
{\bf M.S.} is supported by the program ``Estancias Sab\'aticas para la consolidaci\'on de grupos de investigaci\'on'', SECIHTI.\\ {\bf V.S.} is supported by the program ``Becas Nacionales de Posgrado'', SECIHTI.

\bibliographystyle{unsrt}
\bibliography{biblio}
\end{document}